\documentclass[11pt,oneside]{article}
\usepackage{a4wide}
\usepackage{mathrsfs}
\usepackage{latexsym,bm}
\usepackage{graphicx}
\usepackage{indentfirst}
\usepackage{slashed}
\usepackage{amsmath}
\usepackage{amssymb}
\usepackage{color}
\usepackage[
colorlinks=true,
linkcolor=blue,
breaklinks=true,
urlcolor=magenta,
citecolor=blue]{hyperref}
\usepackage{epsfig}
\usepackage[titletoc]{appendix}
\usepackage{multirow}%
\usepackage{rotating}
\usepackage[square,numbers,sort&compress]{natbib}

\usepackage[normalem]{ulem} 
\usepackage[top=2.9cm,bottom=2.5cm,left=2.8cm,right=3cm]{geometry}
\usepackage{tabularx}

\newcommand{\be}{\begin{equation}} \newcommand{\ee}{\end{equation}}
\newcommand{\ba}{\left(\begin{array}{c}}
\newcommand{\ea}{\end{array}\right)}
\newcommand{\bea}{\begin{eqnarray}} \newcommand{\eea}{\end{eqnarray}}

\newcommand{\zc}{Z_c(3900)}

\newcommand{\dv}{D^*}
\newcommand{\dvb}{\bar{D}^*}
\newcommand{\jpsi}{J/\psi}

\newcommand{\email}[1]{\footnote{\texttt{#1}}}

\newcommand{\bma}{\left(\begin{matrix}}
\newcommand{\ema}{\end{matrix}\right)}
\newcommand{\bqa}{\begin{eqnarray}}
\newcommand{\eqa}{\end{eqnarray}}
\newcommand{\bqaa}{\begin{eqnarray*}}
\newcommand{\eqaa}{\end{eqnarray*}}
\newcommand{\mL}{\mathcal{L}}

\begin{document}
\thispagestyle{empty}
\title{
\Large \bf 
Reconciling experimental and lattice data of $Z_c(3900)$ in a $J/\psi\pi$-$D\bar{D}^*$ coupled-channel analysis
}
\author{\small Lin-Wan Yan$^a$,\, Zhi-Hui Guo$^{a,b}$\email{zhguo@hebtu.edu.cn},\,  Feng-Kun Guo$^{b,c,d}$\email{fkguo@itp.ac.cn},\, De-Liang Yao$^{e,f}$\email{yaodeliang@hnu.edu.cn},\,  Zhi-Yong Zhou$^{g}$\email{zhouzhy@seu.edu.cn} \\[0.5em]
 { \small\it ${}^a$  Department of Physics and Hebei Key Laboratory of Photophysics Research and Application, } \\
 {\small\it Hebei Normal University,  Shijiazhuang 050024, China} \\[0.3em]
{ \small\it ${}^b$ CAS Key Laboratory of Theoretical Physics, Institute of Theoretical Physics, } \\
{\small \it Chinese Academy of Sciences, Beijing 100190, China}\\[0.3em]
{\small \it  $^c$  School of Physical Science, University of Chinese Academy of Sciences, Beijing 100049, China}   \\[0.3em] 
{\small \it  $^d$  Peng Huanwu Collaborative Center for Research and Education, } \\ 
{\small \it Beihang University, Beijing 100191, China} \\[0.3em] 
{\small \it  $^e$  School of Physics and Electronics, Hunan University, Changsha 410082, China} \\[0.3em] 
{\small \it  $^f$  Hunan Provincial Key Laboratory of High-Energy Scale Physics and Applications, } \\ 
{\small \it Hunan University, 410082 Changsha, China} \\[0.3em] 
{ \small\it ${}^g$  School of Physics, Southeast University, Nanjing 211189, China} \\ 
[0.3em]
}
\date{}
  
%

\maketitle
\begin{abstract}
We study the $J/\psi \pi$ and $D\bar{D}^*$ coupled-channel system within a covariant framework. The $J/\psi \pi$ and $D\bar{D}^*$ invariant-mass distributions measured at 4.23~GeV and 4.26~GeV by BESIII and the finite-volume energy levels from recent lattice QCD simulations are simultaneously fitted. Phase shifts and inelasticities of the $J/\psi \pi$ and $D\bar{D}^*$ scattering are predicted using the resulting amplitudes. Poles corresponding to the $Z_c(3900)$ state are found in the complex energy plane and their couplings with $J/\psi \pi$ and $D\bar{D}^*$ are determined.  
Our results indicate that the current lattice data do not preclude the existence of a physical $Z_c(3900)$ state. 
\end{abstract}

\section{Introduction}

The first undoubted tetraquark candidate $\zc$ in the charm sector, which was observed by BESIII and Belle in the $\jpsi\pi^{\pm}$ distributions from the $e^+e^-\to\jpsi\pi^+\pi^-$ process at $\sqrt{s}=4.26$~GeV, has attracted much attention from the experimental, theoretical and lattice QCD communities since its discovery in 2013~\cite{Ablikim:2013mio,Liu:2013dau}. This charged charmoniumlike state was confirmed in an analysis of the CLEO-c data of $e^+e^-\to\jpsi\pi^+\pi^-$ at $\sqrt{s}=4.17$~GeV~\cite{Xiao:2013iha} and in the semi-inclusive decays of $b$-flavored hadrons with $\jpsi\pi^+\pi^-$ in the range 4.2--4.7~GeV by D0 Collaboration~\cite{D0:2018wyb}. The neutral partner of $\zc$ was discovered later by BESIII in the $e^+e^-\to\jpsi\pi^0\pi^0$ process, confirming the $\zc$ as an isovector  state~\cite{Ablikim:2015tbp}. Its spin and parity were unambiguously determined to be $J^P=1^+$ in Ref.~\cite{BESIII:2017bua}. Interestingly similar exotic charmoniumlike $Z_c(3885)$ states, both charged~\cite{Ablikim:2013xfr,Ablikim:2015swa} and neutral~\cite{Ablikim:2015gda} ones, were observed in the $(D\dvb)^{\pm,0}$ distributions in the $e^+e^-\to \pi^{\pm}(D\dvb)^{\mp}$ and $\pi^0(D\dvb)^{0}$ processes. Recent years have witnessed the intensive and interesting  studies on the theoretical explanations of the intriguing $Z_c$ states, including the compact tetraquark  states~\cite{Faccini:2013lda,Dias:2013xfa,Qiao:2013raa,Braaten:2013boa,Wang:2013vex}, 
the kinematical singularities as either threshold cusp  effects~\cite{Chen:2013coa,Liu:2013vfa,Swanson:2014tra,Ikeda:2016zwx} or triangle singularities~\cite{Liu:2015taa,Szczepaniak:2015eza} and the hadronic molecules~\cite{Wang:2013cya,Guo:2013sya,Zhang:2013aoa,Dong:2013iqa,Wilbring:2013cha,Cui:2013yva,Patel:2014vua,Chen:2015ata,Zhou:2015jta,Albaladejo:2015lob,Gong:2016hlt,He:2017lhy,Du:2022jjv}. 

Although it is not definitely clear whether the two resonant structures observed in the $\jpsi\pi$ and $D\dvb$ distributions have the same origin, it is natural to assume that $\zc$ and $Z_c(3885)$, which will be simply denoted as $\zc$ hereafter, correspond to the same state, due to the close values of their masses and widths~\cite{BESIII:2017bua,Ablikim:2013xfr,Ablikim:2015gda,Ablikim:2015swa}. In order to concretely verify this presumption, it is necessary to perform a coupled-channel calculation to simultaneously describe the available experimental data on the $\jpsi\pi$ and $D\dvb$ distributions. It is advocated in Refs.~\cite{Ikeda:2016zwx,Ortega:2018cnm} that the off-diagonal $\jpsi\pi$-$D\dvb$ interaction is mostly responsible for the resonant structure of the $\zc$ (the $\eta_c\rho$-$D\bar D^*$ interaction is also suggested to be important in Ref.~\cite{Ikeda:2016zwx}). 
The importance of triangle singularities in producing the $Z_c(3900)$ signal has been pointed out in Ref.~\cite{Wang:2013cya} immediately after the discovery, and confirmed in Ref.~\cite{Albaladejo:2015lob,Pilloni:2016obd,Du:2022jjv}.
Nevertheless, it is concluded in Refs.~\cite{Wang:2013cya,Albaladejo:2015lob,Du:2022jjv}, which consider both triangle singularities and final state interactions with $J/\psi\pi$-$D\bar D^*$ coupled-channel amplitudes, that a $Z_c(3900)$ pole is still present in the best fits.
It is further shown in Refs.~\cite{Albaladejo:2015lob,Du:2022jjv} that if only  constant contact terms are considered in the perturbative amplitudes for the $J/\psi\pi$-$D\bar D^*$ coupled channels, one can only get a virtual-state like pole for the $Z_c(3900)$, {\it i.e.}, a pole below the $D\bar D^*$ threshold on the unphysical Riemann sheet of the complex energy plane.
If a more general form of the perturbative amplitudes by including an energy-dependent term is considered, the best fit to the BESIII data leads to a resonance pole above the $D\bar D^*$ threshold~\cite{Albaladejo:2015lob,Du:2022jjv}. 
Therefore, it would be important to include general terms in the $\jpsi\pi$ and $D\dvb$ coupled-channel scattering amplitudes to check the robustness.

Moreover, important progresses on the lattice QCD simulations of the $\zc$ have been made by several groups in Refs.~\cite{Chen:2014afa,Prelovsek:2014swa,Ikeda:2016zwx,Ikeda:2017mee,Cheung:2017tnt,Chen:2019iux}. 
Weakly repulsive interaction was revealed for the $D\dvb$ system in a single-channel lattice simulation~\cite{Chen:2014afa}. Later on improvement by considering the coupled-channel $\jpsi\pi$ and $D\dvb$ scattering has been achieved by the CLQCD Collaboration~\cite{Chen:2019iux}. 
Several precise finite-volume energy levels in the center-of-mass (CM) frame were obtained. By using the coupled-channel L\"uscher formalism to fit the energy levels,  no resonant peaks were found near the $D\dvb$ threshold in the $I^{G}(J^{PC})=1^+{(1^{+-})}$ channel~\cite{Chen:2019iux}. Similar conclusions were also obtained earlier in Ref.~\cite{Prelovsek:2014swa} and later by the Hadron Spectrum Collaboration (HSC)~\cite{Cheung:2017tnt}. The HALQCD collaboration performed a three-channel ($\jpsi\pi, \rho\eta_c$ and $D\dvb$) simulation, and found that the $\zc$ could correspond to a threshold cusp~\cite{Ikeda:2016zwx,Ikeda:2017mee}.\footnote{A pole was found far away from the $D\bar D^*$ threshold~\cite{Ikeda:2017mee}.}  

One of the novelties in the present work is to perform a joint analysis of the experimental $\jpsi\pi$ and $D\dvb$ distributions and the lattice finite-volume energy levels resulting from the coupled-channel simulations with $\jpsi\pi$ and $D\dvb$ in Refs.~\cite{Cheung:2017tnt,Chen:2019iux}, within the unitarized partial-wave amplitude approach. In this way we expect to tightly constrain the unknown parameters in the $\jpsi\pi$ and $D\dvb$ scattering amplitudes and draw more definite conclusions on the properties of the $\zc$. 

In this work we use the effective Lagrangian approach to calculate the relevant perturbative amplitudes. Different from the nonrelativistic treatment of the amplitudes~\cite{Aceti:2014uea,Albaladejo:2015lob,Du:2022jjv}, we utilize the manifestly relativistic formalism to perform the partial-wave projections of the coupled $\jpsi\pi$ and $D\dvb$ scattering amplitudes. By taking into account the unitarity conditions, the partial-wave amplitudes are employed to construct the unitarized $T$ matrix, which then provides the key input to analyze the experimental $\jpsi\pi$ and $D\dvb$ event distributions and the lattice finite-volume energy levels.  

This article is organized as follows. In Sec.~\ref{sec.theo}, we introduce the theoretical formalism of the covariant partial-wave amplitude and its unitarization. The inclusion of the finite-volume effects in the unitarized amplitude is also elaborated. In Sec.~\ref{sec.fit}, we perform the fits to the experimental event distributions and the lattice discrete energies. The resulting resonance poles, their couplings and the compositeness are then discussed in detail in Sec.~\ref{sec.pole}. Finally in Sec.~\ref{sec.conclusion} we give a short summary and conclusions.  

\section{Covariant partial-wave amplitude and its unitarization}\label{sec.theo}

The effective Lagrangian that includes the contact $D\dvb \to D\dvb$ vertex reads~\cite{AlFiky:2005jd} 
\begin{align}\label{eq.lagdddvdv}
 \mL_{D\dvb D\dvb}=& -C_{0a} ( D^{\dagger}D + D^{*\dagger}_{\mu} D^{*\mu} ) ( \bar{D}\bar{D}^{\dagger} + \bar{D}^{*}_{\mu} \bar{D}^{*\mu\dagger} )  \nonumber \\ &+ C_{0b} ( D^{\dagger}D^{*}_{\mu} + D^{*\dagger}_{\mu} D ) ( \bar{D}^{*\mu}\bar{D}^{\dagger} + \bar{D} \bar{D}^{*\mu\dagger} )\,,
\end{align}
with 
\begin{eqnarray}
D =   \bma
      D^{+} \\
       D^{0}    \ema\,, 
 \quad D^{\dagger}=(D^{+,\dagger},\,D^{0,\dagger})\,, \quad  \bar{D}=(\bar{D}^{0},\,D^{-})\,, 
 \quad  \bar{D}^{\dagger} =  \bma
       \bar{D}^{0,\dagger} \\
       D^{-,\dagger}   \ema\,. 
\end{eqnarray}
The flavor contents for the vector charmed mesons $D^{*}$ and $\bar{D}^{*}$ are the same as those of the $D$ and $\bar{D}$, respectively. Since we only focus on the $\zc$ channel with definite quantum numbers $I^{G}(J^{PC})=1^+{(1^{+-})}$, only one linear combination of $C_{0a}$ and $C_{0b}$ in Eq.~\eqref{eq.lagdddvdv} will be relevant~\cite{AlFiky:2005jd,Valderrama:2012jv,Guo:2013sya}, which will be simply denoted as $\hat{\lambda}_1$.

We include the following effective operators to describe the interactions between the $\jpsi\pi$ and $D\dvb$ states~\cite{Gong:2016hlt}
\begin{align}\label{eq.lagddvpsipi}
 \mL_{D\dvb\jpsi\pi} =&\hat{\lambda}_2\psi_\mu ( \nabla^{\mu}D^{\dagger} u_\nu \bar{D}^{*\nu\dagger}+\bar{D}^{*\nu}u_\nu \nabla^{\mu}D ) + \hat{\lambda}_3\psi_\mu (\nabla^{\nu}D^{\dagger} u^\mu \bar{D}^{*\dagger}_{\nu} +\bar{D}^{*}_{\nu}  u^\mu \nabla^{\nu}D)
 \nonumber \\ &+ \hat{\lambda}_4 \psi_\mu (\nabla^{\nu}D^{\dagger} u_\nu  \bar{D}^{*\mu\dagger} + \bar{D}^{*\mu} u_\nu  \nabla^{\nu}D  ) +  \hat{\lambda}_5 \psi_\mu (D^{\dagger} \nabla^{\mu}u^\nu  \bar{D}^{*\dagger}_\nu + \bar{D}^{*}_\nu \nabla^{\mu}u^\nu D  )  \,, 
\end{align}
with 
\begin{align}
& \nabla_\nu D^{\dagger} = D^{\dagger}(\overset{\leftarrow}{\partial_\nu}+\Gamma_\nu)\ ,
\quad \nabla_\nu  \bar{D}^{\dagger}=(\partial_\nu+\Gamma_\nu^{\dagger}) \bar{D}^{\dagger}\,, \quad \Gamma_\nu = \frac{1}{2}\bigg( u^\dagger \partial_\nu u+ u \partial_\nu  u^\dagger \bigg)\,, \nonumber \\
& u= e^{i\frac{\Phi}{\sqrt{2}F}}\,,\quad u_\nu =i ( u^\dagger \partial_\nu   u\, -\, u \partial_\nu   u^\dagger )\,, \quad
\Phi =   \bma
      \frac{1}{\sqrt{2}}\pi^0   & \pi^+  \\
      \pi^- & -\frac{1}{\sqrt{2}}\pi^0 \\
   \ema\,.
\end{align}
Next it is straightforward to calculate the scattering amplitudes in the physical bases using the Lagrangians~\eqref{eq.lagdddvdv} and \eqref{eq.lagddvpsipi}, and then transform them into the amplitudes with the proper isospin and $J^{PC}$ quantum numbers that are consistent with the $\zc$ state. In practice, since only the $\zc$ channel with isospin one and $J^{PC}=1^{+-}$ is focused in the present work, we will simply write the amplitudes with definite isospin and $J^{PC}$ in terms of $\lambda_{i=1,\cdots,5}$, which are proportional to the parameters  $\hat{\lambda}_1$ introduced above and $\hat{\lambda}_{i=2,\cdots,5}$ in Eq.~\eqref{eq.lagddvpsipi} in order.  

The $\bar{D}^{*}(a)\,D(b)\to \bar{D}^{*}(c)\,D(d)$ amplitude is given by 
\begin{eqnarray}\label{eq.vfull22}
V_{\dvb D\to \dvb D} = \lambda_1 \varepsilon_c^{\dagger}\cdot\varepsilon_{a}\,.
\end{eqnarray}
The $\jpsi(a)\,\pi(b)\to \bar{D}^{*}(c)\,D(d)$ transition amplitude takes the form 
\begin{eqnarray}\label{eq.vfull12}
 V_{\jpsi\pi\to \dvb D} = \frac{\sqrt{2}}{F_\pi} \bigg(  \lambda_2\, \varepsilon_a\cdot p_d \,\varepsilon_c^{\dagger}\cdot p_b + \lambda_3\, \varepsilon_a\cdot p_b \,\varepsilon_c^{\dagger}\cdot p_d + \lambda_4\,\varepsilon_c^\dagger\cdot\varepsilon_a\,p_b\cdot p_d + \lambda_5\,\varepsilon_a\cdot p_b \,\varepsilon_c^{\dagger}\cdot p_b \bigg)\,.
\end{eqnarray}

In order to implement the unitarity conditions, it is convenient to work with the partial-wave amplitudes. For the scattering processes involving particles with spins, both the $\ell S$ and helicity bases can be used to perform the partial-wave projections. Although in general cases the two approaches are equivalent, the $\ell S$ basis is more suitable for the present study. This is because the $S$-wave interaction of the $D\dvb$ should be the dominant one in the molecular description of the $\zc$, while the $D$-wave part should be much suppressed, at least in the focused energy region around the $D\dvb$ threshold. We follow Ref.~\cite{Gulmez:2016scm} to perform the partial-wave projections in a covariant manner, which improves the nonrelativistic descriptions adopted in Refs.~\cite{Aceti:2014uea,Albaladejo:2015lob,Du:2022jjv}. The covariant approach of the partial-wave projections will automatically introduce certain energy dependent terms to the scattering amplitudes from the polarization vectors, which are of higher orders in the nonrelativistic expansion, without including additional free parameters. 
It is pointed out in Ref.~\cite{Albaladejo:2015lob} that the energy dependence in the interacting kernel is crucial for generating resonance poles of the $\zc$ on the Riemann sheet of the complex energy plane close to the physical region. 
Therefore, we consider that the covariant improvement of the scattering amplitudes should play relevant roles to get further insights into the properties of the $\zc$ despite that the energy dependence is not complete at a given order in the nonrelativistic expansion. 
We give details of the partial-wave projections in Appendix~\ref{sec.appendix}. We will focus on the $S$-wave scattering and the superscript $J=0$ will be omitted throughout for simplicity.

The expressions of the $S$-wave $\jpsi\pi$ (labeled as channel 1) and $\bar{D}^*D$ (labeled as channel 2) scattering amplitudes from Eqs.~\eqref{eq.vfull22} and \eqref{eq.vfull12} take the form 
\begin{align}
 V_{11}(s) = &\, 0\,, \nonumber \\
 V_{12}(s) = &\,  \frac{\sqrt{2}}{9F_\pi M_{D^*}M_{\jpsi}} \bigg\{ \lambda_2\big[ q_2^2(2M_{\jpsi}+E_{\jpsi})E_\pi + q_1^2(2M_{D^{*}}+E_{D^\ast})E_D \big]  \nonumber \\ & 
 -\lambda_4\big[q_1^2 q_2^2 + (2M_{\jpsi}+E_{\jpsi})(2M_{D^{*}}+E_{D^\ast}) E_\pi E_D \big]  
 +\lambda_5\big[q_1^2 \sqrt{s} (2M_{D^{*}}+E_{D^\ast}) \big]  
 \bigg\}\,, \nonumber \\
 V_{22}(s) = & -\frac{\lambda_1}{9M_{D^*}^2}(2M_{D^{*}}+E_{D^\ast})^2 \,, \label{eq.vij}
\end{align}
where the explicit forms of the three-momenta $q_i$ and the energies $E_i$ in the CM frame are 
\begin{align}
 q_1(s) &= \frac{\sqrt{[s-(M_{\jpsi}+M_{\pi})^2][s-(M_{\jpsi}-M_{\pi})^2]}}{2\sqrt{s}}\,,\nonumber \\ 
q_2(s) &= \frac{\sqrt{[s-(M_{D^*}+M_{D})^2][s-(M_{D^*}-M_{D})^2]}}{2\sqrt{s}}\,, \nonumber \\
E_{\jpsi}(s) &= \frac{s+M_{\jpsi}^2-M_{\pi}^2}{2\sqrt{s}} \,, \qquad\qquad 
E_\pi(s) = \frac{s+M_{\pi}^2-M_{\jpsi}^2}{2\sqrt{s}} \,, \nonumber \\ 
E_{D^\ast}(s) &= \frac{s+M_{D^*}^2-M_{D}^2}{2\sqrt{s}} \,,  \qquad \qquad
E_D(s) = \frac{s+M_{D}^2-M_{D^*}^2}{2\sqrt{s}} \,.
\end{align}
Our choice to set $V_{11}=0$, {\it i.e.}, assuming that the perturbative $\jpsi\pi\to\jpsi\pi$ transition amplitude is negligibly small, reconciles with the tiny scattering length of the $\jpsi\pi$ interaction found in Refs.~\cite{Yokokawa:2006td,Liu:2008rza,Liu:2012dv}. It is noted that the $\hat{\lambda}_3$ term in Eq.~\eqref{eq.lagddvpsipi} does not contribute to the $S$-wave amplitude.

The on-shell unitary partial-wave two-body scattering amplitudes can be written as 
\begin{eqnarray}\label{eq.ut}
T(s)= \left[ 1 - N(s)\cdot G(s) \right]^{-1} \cdot N(s)\,, 
\end{eqnarray}
where in the coupled-channel scattering case $T(s)$, $N(s)$ and $G(s)$ should be understood as matrices  spanned in the scattering-channel space. The matrix $N(s)$ here takes the form
\begin{eqnarray}
N(s) =   \bma
      V_{11}(s)  & V_{12}(s)  \\
      V_{12}(s)  & V_{22}(s)  \\
   \ema\,,
\end{eqnarray}
where the matrix elements are given in Eq.~\eqref{eq.vij}. The  $G(s)$ matrix responsible for the right-hand cut is diagonal, $G(s) = \operatorname{diag}(G_1(s), G_2(s))$. The $s$-channel unitarity determines the imaginary part of $G_i(s)$ as
\begin{equation}
 {\rm Im}\, G_i(s) = \frac{q_i(s)}{8\pi\,\sqrt{s}}\,, \qquad (s>s_{{\rm th},i})\,,
\end{equation}
where $s_{{\rm th},i}$ denotes the threshold of the $i$-th channel. Evaluating the $G_i(s)$ function by using the once-subtracted dispersion relation or dimensional regularization, one obtains~\cite{Oller:1998zr} 
\begin{align}\label{eq.gfunc}
G_i(s)  &= -\frac{1}{16\pi^2}\left[ a_{\text{SC},i}(\mu) + \log\frac{m_2^2}{\mu^2}-x_+\log\frac{x_+-1}{x_+}
-x_-\log\frac{x_--1}{x_-} \right]\,, \nonumber\\
 x_\pm &=\frac{s+m_1^2-m_2^2}{2s}\pm \frac{q(s)}{\sqrt{s}}\,,
\end{align}
where $m_1$ and $m_2$ are the masses of the two particles in the considered channel, $\mu$ is an energy scale and $a_\text{SC}(\mu)$ is the subtraction constant. The $G_i(s)$ function has only one free parameter---a change of $\mu$ can be absorbed by a corresponding change of $a_{\text{SC},i}(\mu)$. To be specific, we shall take $\mu=770$~MeV throughout. 

To describe the experimental event distributions, one needs to further consider the production amplitudes, which should incorporate the final-state interactions. Consistent with the recipe of the construction of the unitarized scattering amplitude in Eq.~\eqref{eq.ut}, a similar two-body production formula 
\begin{eqnarray}\label{eq.up}
 \mathcal{P}(s) = \bma
      P_1(s)   \\
      P_2(s)     \\
   \ema = \left[ 1 - N(s)\cdot G(s) \right]^{-1}\cdot \alpha\,,
\end{eqnarray}
with constant production vertices 
\begin{equation}\label{eq.alpha}
 \alpha= \bma
      \alpha_1   \\
      \alpha_2     \\
   \ema\,,
\end{equation} 
has been demonstrated to be able to successfully describe various event distributions~\cite{Oller:2000fj,Meissner:2000bc,Guo:2011pa,Guo:2016zep,Guo:2016bjq,Guo:2020pvt}. 
To be more specific, the functions $P_1(s)$ and $P_2(s)$ describe the $\jpsi\pi$ and $D\dvb$  production amplitudes, respectively. In this way, once the unknown parameters in the production amplitudes~\eqref{eq.up} are determined through fitting to data, the unitarized scattering amplitudes of Eq.~\eqref{eq.ut} will be totally fixed. The resonance information of the $\zc$, including its pole positions and coupling strengths to the considered channels, can then be extracted from the unitarized scattering amplitudes. In fact, all the unknown parameters describing the discrete lattice energy levels that will be discussed later also appear in the production amplitudes in Eq.~\eqref{eq.up}. Therefore, this enables us to perform a joint fit to both the experimental and lattice data. 

The experimental event distributions of the $\jpsi\pi$ and $D\dvb$ channels are projected from the three-body decays $Y\to \jpsi\pi\pi$ and $D\dvb\pi$, respectively. To account for the strong  coupled-channel final-state interactions, we use the  production amplitudes in Eq.~\eqref{eq.up} to  construct the full decay amplitude for the $Y\to \jpsi\pi\pi$ process  
\begin{eqnarray}\label{eq.decayamp1p}
 M_1(s,t)= \epsilon_Y^{\dagger} \cdot \epsilon_{\jpsi} \big[ P_1(s) + P_1(t)  \big] \,,
\end{eqnarray}
where $\epsilon_Y^{\dagger}$ and  $\epsilon_{\jpsi}$ stand for the polarization vectors, $s$ and $t$ correspond to the invariant energy squared of the $\jpsi\pi$ systems. To reasonably reproduce the experimental $\jpsi\pi$ line shapes, it is necessary to include the background contributions, such as the effects from the crossed $\pi\pi$ channels~\cite{Gong:2016hlt,Chen:2019mgp,Pilloni:2016obd}. Following the recipes from Refs.~\cite{Ablikim:2013mio,Albaladejo:2015lob}, we parameterize the background effects in the $\jpsi\pi$ event distribution as 
\begin{eqnarray}\label{eq.bk}
B_1= b_1[(\sqrt{s}-M_{\jpsi}-M_\pi)(M_Y-M_\pi-\sqrt{s})]^{c_1}\,,
\end{eqnarray} 
where the unknown parameters $b_1$ and $c_1$ will be determined by data.  
For the $Y\to D\dvb\pi$ process, the experimental double-$D$ tag data indicate that the background contributions to the $D\dvb$ invariant-mass distributions are tiny~
\cite{Ablikim:2015swa}. We will simply subtract the background events from the BESIII analysis out from the $D\dvb$ event distribution. Then we fit to the resulting $D\dvb$ event distribution with a vanishing background term $B_2(s)=0$.
The $Y\to D\dvb\pi$ decay amplitude reads
\begin{eqnarray}\label{eq.decayamp2p}
 M_2(s,t)= \epsilon_Y^{\dagger} \cdot \epsilon_{\dv}   P_2(s) \,,
\end{eqnarray}
with $s$ and $t$ the energy squared of the $D\dvb$ and $D\pi$ systems, respectively. 
The experimental event distributions are fitted using
\begin{equation}\label{eq.distn}
 \frac{d N_{i}}{d\sqrt{s}}= A_i(s) + B_i(s) \,,
\end{equation} 
with 
\begin{eqnarray}\label{eq.dista}
A_i(s)=\int_{t_{i,-}}^{t_{i,+}}\frac{1}{(2\pi)^3}\frac{1}{32 M_Y^3}\, \frac{1}{3}\sum_{spins} \big| M_i(s,t) \big|^2 dt\,,
\end{eqnarray}
where $t_{i,-}$ and $t_{i,+}$ stand for the kinematic boundaries of the $i$-th process, being $i=1$ for $Y\to\jpsi \pi\pi$ and $i=2$ for $Y\to D\dvb\pi$.

In order to describe the lattice energy levels, we need to put the unitarized amplitudes into a finite volume. We will utilize the method proposed in Refs.~\cite{Doring:2011vk,Doring:2012eu}, which has been successfully applied to fit the lattice energy levels in many processes, such as the $\pi\eta$-$K\bar{K}$-$\pi\eta'$ coupled-channel scattering process~\cite{Guo:2016zep} and the $D\pi$-$D\eta$-$D_s\bar{K}$ coupled-channel scattering~\cite{Guo:2018tjx}. In this formalism, the finite-volume effects are introduced via the $G(s)$ function, while finite-volume corrections to the tree-level partial-wave amplitudes in Eq.~\eqref{eq.vij}, happening at shorter distances, are exponentially suppressed and thus neglected. For a cubic box of length $L$ with periodical boundary conditions, the finite-volume correction of the $G(s)$ function in the CM frame is given by~\cite{Doring:2011vk,Doring:2012eu,Guo:2016zep,Guo:2018tjx}
\begin{eqnarray}\label{eq.deltag}
\Delta G(s) &=& \frac{1}{L^3} \sum_{\vec{n}}^{|\vec{q}|<q_{\rm max}} I(|\vec{q}|) -   \int^{|\vec{q}|<q_{\rm max}} 
\frac{{\rm d}^3 \vec{q}}{(2\pi)^3} I(|\vec{q}|)\,, 
\end{eqnarray}
with 
\begin{eqnarray}\label{eq.fvkin}
&& \vec{q}= \frac{2\pi}{L} \vec{n},\,(\vec{n} \in \mathbb{Z}^3) \,, \quad
 I(|\vec{q}|) = \frac{\omega_1+\omega_2}{2\omega_1 \omega_2 \,[s-(\omega_1+\omega_2)^2]}\,, \quad
\omega_i =\sqrt{|\vec{q}|^2+m_i^2}  \,. 
\end{eqnarray} 
We introduce the three-momentum cutoff $q_{\rm max}$ to make the discrete sum and the continuous integral convergent in Eq.~\eqref{eq.deltag}. 
Since the finite-volume corrections are of long distance nature, $\Delta G(s)$ depends little on the ultraviolet regulator, which is the hard cutoff $q_\text{max}$, as has been explicitly demonstrated in, {\it e.g.}, Ref.~\cite{Guo:2016zep}. 
Then in the finite-volume study the function $G(s)$ in Eq.~\eqref{eq.ut} should be replaced by  
\begin{eqnarray}\label{eq.gfuncfvdr}
\widetilde{G}(s)= G(s) + \Delta G(s) \,,
\end{eqnarray}
where $G(s)$ and $\Delta G(s)$ are given in Eqs.~\eqref{eq.gfunc} and \eqref{eq.deltag}, respectively. The finite-volume spectra can be obtained by solving 
\begin{eqnarray}\label{eq.detfv}
 \det{[1 - N(s) \cdot \widetilde{G}(s)]} =0\,.
\end{eqnarray}
In this work we only analyze the discrete finite-volume data obtained in the CM frame~\cite{Chen:2019iux,Cheung:2017tnt}. In this case it is enough for us to rely on Eq.~\eqref{eq.detfv} to fit the lattice data. For the general case in a moving frame, one should accordingly use different formulas by properly including the possible mixing between different partial waves~\cite{Doring:2012eu,Gockeler:2012yj,Guo:2018tjx}. 
{The dispersion relation between the energy and mass of the charm meson adopted in the lattice QCD simulation of Ref.~\cite{Prelovsek:2014swa} is different from the relativistic one, $\omega_i =\sqrt{|\vec{q}|^2+m_i^2}$, used in Eqs.~\eqref{eq.deltag} and \eqref{eq.fvkin}. 
It has been demonstrated in Ref.~\cite{Albaladejo:2016jsg} that the lattice results in Ref.~\cite{Prelovsek:2014swa} are consistent with the amplitude determined in Ref.~\cite{Albaladejo:2015lob}, which allows for either a virtual state or a resonance $Z_c(3900)$ pole. To be consistent with the relativistic kinematical relations used in our finite-volume setups~\eqref{eq.fvkin}, we will include the lattice results only from Refs.~\cite{Chen:2019iux,Cheung:2017tnt} in later fits. 

\section{Global fits to the experimental and lattice data}\label{sec.fit}

We determine the unknown parameters through simultaneous fits to experimental and lattice data. 
The relevant experimental data include the $\jpsi\pi^{\pm}$ event distributions from the $e^+e^-\to \jpsi\pi^+\pi^-$ process at the $e^+e^-$ CM energies 4.23~GeV and 4.26~GeV~\cite{BESIII:2017bua}, and the $D^0 D^{*-}$ and $D^- D^{*0}$ event distributions from the $e^+e^-\to \pi^\pm(D\dvb)^{\mp}$ processes at the same $e^+e^-$ CM energies~\cite{Ablikim:2015swa}. 
For the lattice data, the finite-volume spectra given in Refs.~\cite{Chen:2019iux,Cheung:2017tnt}, whose simulations are done with unphysically large pion masses but with the charm quark mass close to its physical value, will be analyzed. The masses used in the lattice simulations are $m_\pi=391$~MeV, $M_{D}=1885$~MeV, $M_{D^*}=2009$~MeV, $M_{\jpsi}=3045$~MeV in Ref.~\cite{Cheung:2017tnt}, and $m_\pi=324$~MeV, $M_{D}=1822$~MeV, $M_{D^*}=2029$~MeV, $M_{\jpsi}=2969$~MeV in Ref.~\cite{Chen:2019iux}, which will be directly taken in our fits to the lattice finite-volume energy levels. 

The roles of the triangle diagrams, with $D_1, D^{*}$ and $D$ running in the loops, are currently under vivid discussions for the production of the $\zc$ peaks in the $e^+e^-\to \jpsi\pi\pi$ and $\pi D\dvb$~\cite{Wang:2013cya,Wang:2013kra,Albaladejo:2015lob,Gong:2016jzb,Pilloni:2016obd,Du:2022jjv}. 
It is noticed that the triangle singularities are rather sensitive to the specific $e^+e^-$ energies~\cite{Guo:2019twa}. In order to further check the possible energy sensitivity, we first take the strategy to separately fit the $\jpsi\pi$ and $D\dvb$ event distributions obtained from the two different $e^+e^-$ energies, namely 4.23~GeV and 4.26~GeV. In this way we do not explicitly introduce the triangle diagrams in the production mechanisms, but rather we use the same theoretical formalism to individually fit the data at 4.23~GeV and 4.26~GeV. If the resulting parameters from the two fits resemble each other, it indicates that that the triangle diagrams are not necessarily the exclusive source for the $\zc$. Otherwise, if the resulting parameters from the two different fits are rather distinct, it is quite plausible that the triangle singularities indeed can be decisive in the description of the experimental data for the $\zc$ peaks.

Regarding the background terms~\eqref{eq.bk} in the $\jpsi\pi$ event distributions, it is found that good reproduction of the experimental data can be achieved by releasing $b_1$ and fixing $c_1=1$ in our fits, while in addition to $b_1$ the parameter $c_1$ is also set to free in Refs.~\cite{Albaladejo:2015lob,Ablikim:2013mio}. Due to the obvious differences of the experimental $\jpsi\pi$ event distributions at 4.23~GeV and 4.26~GeV, two different values of $b_1$ will be separately fitted to the data at the two different $e^+e^-$ energy points.

The coupling $\lambda_1$ in Eq.~\eqref{eq.vfull22} is dimensionless. For the couplings $\lambda_{i=2,3,4,5}$ in Eq.~\eqref{eq.vfull12}, they have the mass dimension, and it is convenient to define dimensionless ones as 
\begin{eqnarray}
 \tilde{\lambda}_{i=2,3,4,5}= \frac{\lambda_{i=2,3,4,5}}{M_D}\,. 
\end{eqnarray}
The $\lambda_3$ term in Eq.~\eqref{eq.vfull12} will not contribute after the partial-wave projection. The remaining $\tilde{\lambda}_{2}, \tilde{\lambda}_4$ and $\tilde{\lambda}_5$ will be fitted. The subtraction constants $a_{\rm SC}^{\jpsi\pi}$ and $a_{\rm SC}^{D\dvb}$ introduced through the unitarization procedure are allowed to float in our fits. 
Since the production of open-charm $D\bar D^*$ channel is much easier than that of the $J/\psi\pi$---the charm and anti-charm quark pair produced in $e^+e^-$ collisions need to fly with similar momenta to form $J/\psi$ and thus have a very limited phase space, it is reasonable to set $\alpha_1=0$ for the production  vertices~\eqref{eq.alpha}.\footnote{The cross section of $e^+e^-\to \pi^+D^0D^{*-}$ between 4.2 and 4.3 GeV is about 200--300~pb~\cite{BESIII:2018iea}, while that of $e^+e^-\to \pi^+\pi^-J/\psi$ is several times smaller, about 50--80~pb~\cite{BESIII:2016bnd}.} 
For $\alpha_2$, one needs to separately fit its values in the $\jpsi\pi$ and $D\dvb$ event distributions at 4.23~GeV and 4.26~GeV, in order to account for the various experimental factors for the two different channels at different $e^+e^-$ energy points.

\begin{table}[tb]
\centering
\caption{\label{tab.fitlecs}  Resulting parameters from the fits. For the notations of different fits and parameters, see the text for details. For the joint fit, the left/right numbers for the entries $b_1$ and $\alpha_2$ correspond to the results from the data sets at $4.23$~GeV and $4.26$~GeV, respectively.  }
\begin{tabular}{l c c c c c}
\hline\hline
 & Fit-4230 & Fit--4260  & Joint Fit
\\  \hline
$a_{\text{SC},1}$ &$-3.76^{+0.38}_{-0.47} $&$-4.13^{+0.51}_{-0.71} $  &$-4.02^{+0.32}_{-0.49}$\\
$a_{\text{SC},2}$ &$-2.82^{+0.04}_{-0.03}$  &$-2.82^{+0.03}_{-0.04}$ &$-2.80^{+0.02}_{-0.02}$\\
$\lambda_1$&$ -88^{+18}_{-14}$&$ -63^{+28}_{-21}$& $-86^{+12}_{-11}$\\
$\tilde{\lambda}_2$& $1064^{+103}_{-127}$& $1056^{+112}_{-139}$& $1082^{+\ 93}_{-116}$\\
$\tilde{\lambda}_4$&$ -39^{+15}_{-11}$&$ -40^{+16}_{-12}$& $-41^{+14}_{-11}$\\
$\tilde{\lambda}_5$& $-725^{+149}_{-114}$& $-725^{+164}_{-121}$& $-751^{+137}_{-111}$\\
{$b_1$(MeV$^{-3}$)}& $(8.6^{+0.2}_{-0.2}) \times 10^{-4}$& $(4.7^{+0.2}_{-0.1}) \times 10^{-4}$& $(8.7^{+0.2}_{-0.2})/(4.8^{+0.1}_{-0.2}) \times 10^{-4}$\\
{$|\alpha^{J/\psi \pi}_2|^2$}&$17.5^{+1.8}_{-1.1}$&$9.1^{+2.3}_{-1.5}$&$19.4^{+3.4}_{-2.1}/9.0^{+1.6}_{-1.1}$\\
{$|\alpha^{D\bar{D}^* }_2|^2$}&$2.9^{+0.7}_{-0.5}$&$1.5^{+0.6}_{-0.4}$&$2.7^{+0.6}_{-0.6}/1.3^{+0.3}_{-0.3}$\\
${\chi^2}/{\text{d.o.f}}$&$1.31$&$1.16$&$1.18$\\
\hline\hline
\end{tabular}
\end{table}

The resulting parameters from the two separate fits are summarized in the columns labeled as Fit-4230 and Fit-4260 in Table~\ref{tab.fitlecs}. It is clear that the parameters that determine the $\jpsi\pi$ and $D\dvb$ amplitudes from the two fits are perfectly consistent with each other. We verify that the qualities of the reproduction of the experimental and lattice data from the two types of fits are also quite similar, which can be inferred from the close values of $\chi^2$ given in Table~\ref{tab.fitlecs}. This important finding indicates that the finite-state interactions between the $\jpsi\pi$ and $D\dvb$ are able to reasonably describe the $\zc$ peaks observed at different $e^+e^-$ energies. It also supports that  triangle singularities in the $\zc$ production does not seem to play the exclusively decisive role~\cite{Wang:2013cya,Guo:2014iya,Gong:2016jzb,Albaladejo:2015lob,Du:2022jjv}.

\begin{figure}[tbhp]
   \centering
   \includegraphics[width=0.8\textwidth,angle=-0]{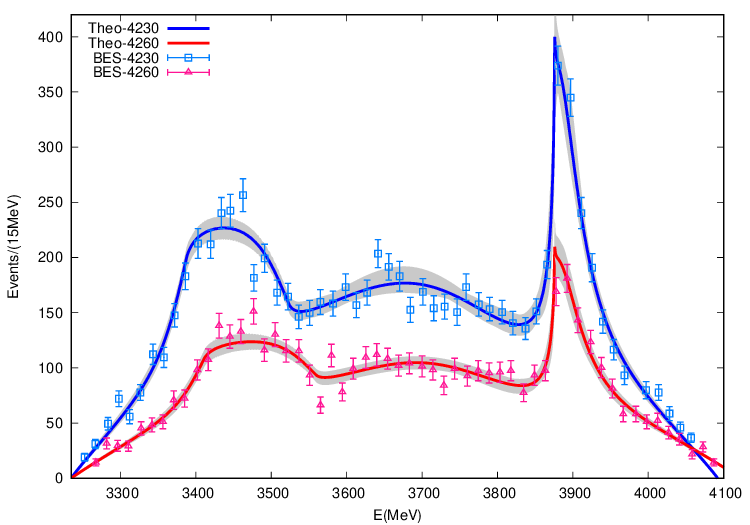} 
   \includegraphics[width=0.8\textwidth,angle=-0]{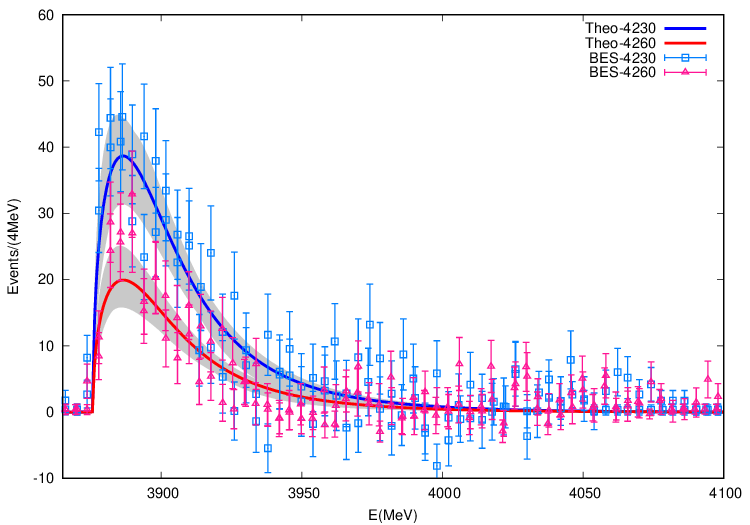} 
  \caption{Joint fit to the $\jpsi\pi$ (upper) and  $D\dvb$ (lower) event distributions from BESIII. The data are taken from Refs.~\cite{BESIII:2017bua} and \cite{Ablikim:2015swa}, respectively. For the $D\dvb$ event distributions, the background events from the experimental analysis~\cite{Ablikim:2015swa} are subtracted. The shaded areas correspond to the uncertainties propagated from fitting using the bootstrap method. }
   \label{fig.fitpsipi}
\end{figure}

\begin{figure}[htb]
     \centering
     \includegraphics[width=0.95\textwidth,angle=-0]{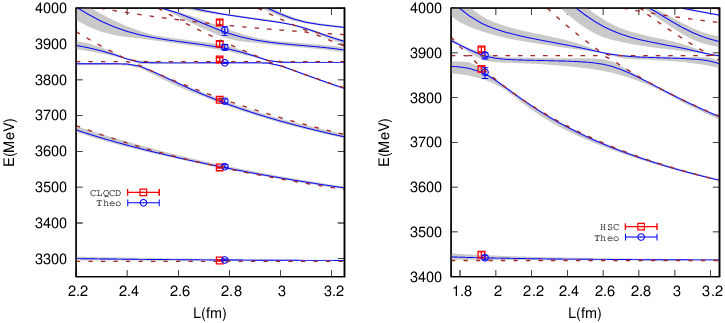} 
    \caption{Finite volume energy levels from the joint fit. For the CLQCD data from Ref.~\cite{Chen:2019iux} we have taken the six energy levels below  4.0~GeV in the fit. The HSC data are taken from Ref.~\cite{Cheung:2017tnt}. The blue solid lines correspond to our theoretical predictions by continuously varying the box length $L$, and the gray bands correspond to the uncertainties propagated from the fit. The brown dashed lines stand for the free energy levels.
    The theoretical results shown as blue circles with error bars are slightly shifted to the right in order to be distinguished from the lattice data shown as red squares.
    }
    \label{fig.fitlat}
\end{figure}

Because of the similarities of the resulting parameters in the scattering amplitudes from the separate fits, it is meaningful to further perform a joint fit to simultaneously include the two sets of experimental data obtained at 4.23~GeV and 4.26~GeV, together with the lattice discrete spectra.  
It is expected that the joint fit can provide more reliable and tighter constraints on the coupled-channel scattering amplitudes of $\jpsi\pi$ and $D\dvb$. 
Reasonably good reproductions of the experimental and lattice data from the joint fit are shown  in Figs.~\ref{fig.fitpsipi} and \ref{fig.fitlat}.
The values of the parameters from the joint fit are given in the last column in Table~\ref{tab.fitlecs}. The two subtraction constants, $\lambda_{1}$ and $\tilde{\lambda}_{2,4,5}$, describing the nonperturbative interactions of the $\jpsi\pi$ and $D\dvb$, enter their coupled-channel scattering amplitudes, which in turn affect the event distributions at all the energy points and also the lattice finite-volume spectra. In other words these six parameters simultaneously influence all the considered experimental and lattice data. In contrast, the parameters $b_1$ in the background and the production vertex $\alpha_{2}$, which acts as a normalization constant, are expected to vary with different data sets. It turns out that the parameters from the joint fit are compatible with the separate fits within uncertainties. This fact shows a clear sign of the stability of the fits shown in Table~\ref{tab.fitlecs}. 

Based on the unitarized decay amplitudes in Eqs.~\eqref{eq.decayamp1p},\eqref{eq.decayamp2p}, we estimate the branching ratios of the $\zc$ decaying into $D\bar{D}^*$ and $\jpsi\pi$ channels, by performing the phase space integrals of Eq.~\eqref{eq.dista} in the energy range of $\sqrt{s}=(3900\pm 35)$~MeV, as proposed in Ref.~\cite{Albaladejo:2015lob}. When evaluating the ratio of the partial decay widths of $\zc$ using Eq.~\eqref{eq.dista}, the background contributions are excluded and the $\alpha_2^{J/\psi\pi}$, $\alpha_2^{ D \bar D^*}$ factors accounting for the normalizations of experimental event distributions are set to unity.
The resulting branching ratios of $R=\Gamma_{\zc\to D\bar{D}^*}/\Gamma_{\zc\to\jpsi\pi}$ from the various fits are 
\begin{eqnarray}
R= 1.6^{+0.4}_{-0.4} ~ \,(\text{Fit-4230})\,, \quad 
R= 1.5^{+1.2}_{-0.8}~ \,(\text{Fit-4260})\,, \quad 
R= 1.9^{+0.9}_{-0.6}~ \,(\text{Joint\,Fit})\,.
\end{eqnarray}
The central values are smaller than that from the experimental determination $R^{\rm Exp}=6.2\pm2.9$~\cite{Ablikim:2013xfr}. Taking into account the large  uncertainties, the difference between the one from our joint fit from the BESIII determination is merely slightly larger than $1\sigma$. 
Notice that we do not consider the $\eta_c\rho$ channel since the statistics of the data in this channel is not high (the statistical significance of $Z_c(3900)$ in $\eta_c\rho$ is $3.9\sigma$~\cite{BESIII:2019rek}), and $R$ may be underestimated in our calculation.

The theoretical uncertainties, shown as shaded areas in Figs.~\ref{fig.fitpsipi} and \ref{fig.fitlat},  are calculated through the bootstrap method. In this procedure, we implement the normal distributions by taking the central values and uncertainties of the experimental and lattice data as the means and variances, respectively, to randomly generate large amounts of pseudo-data sets. The pseudo-data sets are then used to repeat the fits.  The large samples of the parameter configurations from the repeated fits using the random pseudo-data sets are further exploited to obtain the uncertainties of the theoretical quantities. When performing the fits to the $D\dvb$ event distributions from Ref.~\cite{Ablikim:2015swa}, the experimental background effects are subtracted. In Fig.~\ref{fig.fitlat}, we also give the predictions to the lattice discrete spectra in a wide range of the volume sizes, which hopefully can provide useful guidelines for future lattice simulations.

\section{Insights into the $\zc$ resonance poles}\label{sec.pole}

Kinematical singularities, such as the two-body cusps and triangle or box singularities, provide one possible source to explain some structures observed in experiments~\cite{Guo:2019twa}. Such kind of singularities is highly sensitive to the kinematics of the processes. On the contrary, resonance poles in the complex energy plane are universal in all production amplitudes involving the same set of particles.  They are expected to show up in all relevant coupled channels, though not necessarily as peaks~\cite{Dong:2020hxe}. In this case details of the kinematics do not play a crucial role, although they may affect the resonant line shapes. Therefore it is of utmost importance to discern whether there exist relevant resonance poles in the system under study. 

The physical scattering amplitudes in Eq.~\eqref{eq.ut} can be extrapolated to the complex energy plane via the unitarity $G(s)$ functions. The expression in Eq.~\eqref{eq.gfunc} stands for the $G(s)$ on the physical/first Riemann sheet (RS) and its corresponding result on the second RS takes the form~\cite{Oller:1998zr} 
\begin{eqnarray}
 G(s)^{\rm II} = G(s) - i \frac{q(s)}{4\pi\sqrt{s}}\,,
\end{eqnarray}
being $q(s)$ the magnitude of the CM three-momentum. In our convention, the imaginary part of $G(s)$ on the first RS is positive in the energy region above the threshold, and the imaginary part of $G(s)^{\rm II}$ is negative. The scattering amplitudes can be analytically continued to different RSs by taking the proper combinations of the $G(s)$ and $G(s)^{\rm II}$ for different channels. For instance, the amplitudes on the second RS can be labeled as $(-,+)$, where the minus/plus sign at each entry corresponds to taking $G(s)^{\rm II}$/$G(s)$ for that channel. In such convention, the amplitudes on the first, third and fourth RSs are given by $(+,+)$, $(-,-)$ and $(+,-)$, respectively. 

Next we search for resonance poles of the scattering amplitudes in the complex plane. At each pole, the singular term of the Laurent expansion of the scattering amplitude takes the form
\begin{eqnarray}\label{eq.defresidue}
T_{ij}(s) = - \frac{\gamma_i\gamma_j}{s-s_R}\,,
\end{eqnarray}
where the indexes $i,j$ stand for the coupled channels of the considered system, and the resonance pole is given by $\sqrt{s_R}=(M_R-i\Gamma_R/2)$, being $M_R$ and $\Gamma_R$ the mass and width, respectively. 
The relevant resonance pole positions, together with their residues $\gamma_i$, are summarized in Table~\ref{tab.phypole}. Around the $\zc$ region, three types of resonance poles are found by taking into account the uncertainties of the parameters in Table~\ref{tab.fitlecs}. We stress that in fact for each parameter configuration it gives two types of poles: one in the third RS and the other one located on either the second RS or the fourth RS, depending on the specific parameter set taken from the large bootstrap samples in the uncertainty analysis. 
The central values of the parameters in Table~\ref{tab.fitlecs} lead to the fourth sheet pole shown in Table~\ref{tab.phypole}. In fact the upper half plane of the second RS is continuously connected to the lower half plane of the fourth RS in the energy region above the $D\dvb$ threshold. 
Therefore, when the pole positions on the second and fourth sheets are similar, the amplitude with a pole with a small imaginary part near threshold on the second RS is expected to resemble the one with a fourth-sheet nearby pole. 
Besides, other distant poles, including the spurious ones in the first sheet, can be also found in the complex plane on different RSs. 
For instance, a fourth-sheet pole around $(3800-6i)$~MeV is found in the amplitude. These remote poles are far away from the interested energy region and do not seem to visibly affect the physical amplitudes, we do not discuss them any further. 

In addition to the preferred fit presented in Table~\ref{tab.fitlecs}, we also find another type of joint fit, which can reasonably reproduce the experimental and lattice data to some extent with somewhat larger $\chi^2/{\rm d.o.f}\simeq 1.45$ and has only a virtual state pole around 3.8~GeV in the amplitude. It is interesting to point out that our finding here is similar to that in a recent study of Ref.~\cite{Du:2022jjv}: the solutions with resonance poles around the $D\dvb$ threshold can better reproduce the experimental data than the ones with only virtual state poles around 3.8~GeV.

\begin{table}[tbh]
\centering
\caption{\label{tab.phypole} Resonance pole positions and their residues. $\gamma_1$ and $\gamma_2$ correspond to the residues of $\jpsi\pi$ and $D\dvb$, respectively. Large amount of bootstrap samples of parameters are generated in our uncertainty analyses. For each parameter sample, it can give a couple of poles, one on the RS III and the other one either on RS II or RS IV. }
\begin{tabular}{l c c c c c c c c c}
\hline\hline
  RS & $M_R$  (MeV)  & $\Gamma_R$/2 (MeV)   & $|\gamma_{1}|$ (GeV)  & $|\gamma_{2}|$ (GeV)
\\ \hline\hline
III 
& $3874.8^{+3.7}_{-4.0}$& $32.7 ^{+1.6}_{-1.9}$&$4.3^{+0.3}_{-0.3}$&$8.7 ^{+0.8}_{-0.7}$ \\
\hline
II 
&$3902.7^{+1.3}_{-1.3}$  & $3.0 ^{+2.4}_{-2.4}$&$4.9 ^{+0.2}_{-0.2}$&$8.3^{+0.3}_{-0.3}$  
           \\
IV 
& $3902.4 ^{+1.2}_{-2.2}$&$ 3.3^{+4.4}_{-2.6}$&$4.6^{+0.3}_{-0.4}$&$8.6 ^{+0.4}_{-0.3}$
\\ 
\hline\hline
\end{tabular}
\end{table}

According to Morgan's pole counting criteria~\cite{Morgan:1992ge}, an elementary resonance state would correspond to two similar poles near the threshold on different RSs. For the molecular type of resonance state, there would be just one pole near the threshold. The situation for the $\zc$ poles in Table~\ref{tab.phypole} is very subtle. Indeed two poles near the $D\dvb$ threshold are found for each parameter configuration. 
However, a closer look at their positions on different RSs reveals that the two poles are not really so close. The imaginary part of the third-sheet pole is one-order larger in magnitude than the one on the second or fourth sheet. Besides, the real part of the pole on the third RS is below the $D\dvb$ threshold, while the pole on the second or fourth RS lies above that threshold. 
A qualitative conclusion  would be that the $\zc$ corresponds to a state lying between the elementary and molecular types, or in another word both types of  components are likely to be of similar importance in the $\zc$ compositions. 
This conclusion can be further quantitatively verified by using the recipe proposed in Ref.~\cite{Guo:2015daa} to calculate the compositeness coefficients for the resonances, which has been applied in the study of various exotic hadronic states~\cite{Guo:2019kdc,Guo:2020pvt,Guo:2020vmu,Du:2021bgb}. 
The partial compositeness coefficient $X_k$, $i.e.$, the probability to find the two-body component from the $k$th channel in the considered resonance state, is given by~\cite{Guo:2015daa} 
\begin{eqnarray}\label{eq.Xg}
 X_k= |\gamma_k|^2 \left| \frac{d G_k^{(\rm II)}(s_R)}{d s} \right|\,,
\end{eqnarray}
where $\gamma_k$ is the residue defined in Eq.~\eqref{eq.defresidue} at the pole $s_R$. Depending on the RS of the location of the pole, one should take the derivation of the proper $G(s)$ or $G^{\rm II}(s)$ with respect to the $s$ in Eq.~\eqref{eq.Xg}. There is a caveat when applying Eq.~\eqref{eq.Xg} to calculate the compositeness $X$: this recipe can not be generally used for any resonance pole. The working condition for Eq.~\eqref{eq.Xg} is that the resonance pole should lie above the nearby threshold~\cite{Guo:2015daa}. However, since the resonance pole on the third RS in Table~\ref{tab.phypole} is rather close to the $D\dvb$ threshold, it is expected that  Eq.~\eqref{eq.Xg} could be also applied to estimate the compositeness of the third-sheet pole. The compositeness coefficients contributed from the $D\dvb$ channel to the three kinds of resonance poles in Table~\ref{tab.phypole} are found to be\footnote{Since the $\jpsi\pi$ channel is rather distant from the $\zc$ pole position, $X_1$ computed using Eq.~\eqref{eq.Xg} is one order of magnitude smaller than $X_2$.}
\begin{eqnarray}\label{eq.xivalue}
X_2 =0.38^{+0.03}_{-0.03} \,\, ({\rm  RS\,II})\,, \quad  X_2= 0.43^{+0.09}_{-0.06}   \,\, ({\rm RS\,III})\,, \quad  X_2 =0.42^{+0.04}_{-0.03}  \,\, ({\rm  RS\,IV})\,,
\end{eqnarray}
which perfectly agree with the determinations from Ref.~\cite{Guo:2020vmu}. 

Other ways to calculate the compositeness coefficients for resonances have also been proposed based on the scattering length ($a$) and effective range ($r$) in Refs.~\cite{Hyodo:2013iga,Matuschek:2020gqe}. The proposal from Ref.~\cite{Hyodo:2013iga} is   
\begin{eqnarray}\label{eq.defxh}
\bar{X} =\sqrt{\left|\frac{1}{1+2 r/a}\right|}\,,
\end{eqnarray}
and the expression from Ref.~\cite{Matuschek:2020gqe} reads
\begin{eqnarray}\label{eq.defxm}
\hat{X} =\sqrt{\frac{1}{1+|2 r / a|}} \,.    
\end{eqnarray}
By taking the joint fit in Table~\ref{tab.fitlecs}, our predictions to the values of the scattering length and effective range for the $\bar{D}D^*$ channel read 
\begin{eqnarray}
a_{\bar{D}D^*}= \big( -0.56^{+0.11}_{-0.13} -i0.28^{+0.04}_{-0.04}  \big) \,{\rm fm}\,, \qquad r_{\bar{D}D^*}= \big(-2.03^{+0.51}_{-0.68} -i0.77 ^{+0.24}_{-0.13}\big) \,{\rm fm}\,.
\end{eqnarray}
The two proposals in Eqs.~\eqref{eq.defxh} and \eqref{eq.defxm} lead to almost identical results $\bar{X}_{2}=\hat{X}_2=0.36^{+0.06}_{-0.06}$, which are compatible with the values in Eq.~\eqref{eq.xivalue} within uncertainties. 

The above results of the $Z_c$ compositeness indicate
that the energy dependence in the contact amplitudes \eqref{eq.vij} (and thus interaction of finite range) plays an important role in forming the $Z_c$ pole of this analysis. 
To probe the origin of the energy dependence, from either other degrees of freedom with higher masses or compact quark state cores, one needs to rely on  the specific theoretical models, which is beyond the scope of our present study. 

\begin{figure}[tbh]
 \centering
 \includegraphics[width=0.99\textwidth,angle=-0]{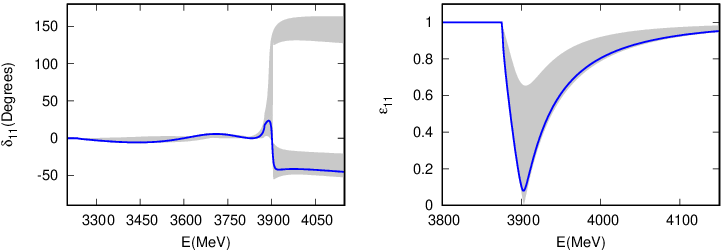} 
\caption{ Phase shifts and inelasticities of the $\jpsi\pi\to\jpsi\pi$ scattering. }
 \label{fig.del11}
\end{figure} 

\begin{figure}[tbh]
 \centering
 \includegraphics[width=0.99\textwidth,angle=-0]{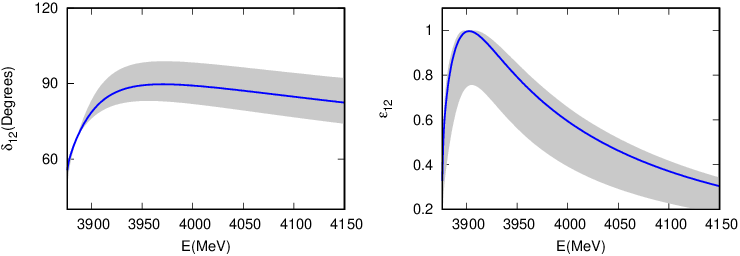} 
\caption{ Phase shifts and inelasticities of the $\jpsi\pi\to D\dvb$ scattering. }
 \label{fig.del22}
\end{figure}

The pole singularities on the unphysical RSs in the complex energy plane also manifest themselves in the physical amplitudes, which can be characterized by the phase shifts and inelasticities in the coupled-channel scattering processes. 
In Figs.~\ref{fig.del11} and  \ref{fig.del22}, we give our predictions of the phase shifts and inelasticities, together with the theoretical uncertainties, for the $\jpsi\pi\to\jpsi\pi$ and $\jpsi\pi\to D\dvb$ processes, respectively. Within the uncertainties there are two branches of phase shifts for the $\jpsi\pi\to\jpsi\pi$ above the $D\dvb$ threshold, and we confirm that the lower branch corresponds to the parameter samples with a pole on the fourth RS and the upper one corresponds to the parameter samples with a pole on the second RS. 
This also tells us that the narrow poles on the second and fourth sheets in Table~\ref{tab.phypole} are the most responsible ones for the $\zc$ signals in our amplitudes.  The two phase-shift branches in Fig.~\ref{fig.del11} differ by about 180 degrees above the $D\dvb$ threshold. As a result they actually lead to rather similar $S$ matrix elements. 
 
\section{Summary and conclusions}\label{sec.conclusion}

In this work we calculate the covariant partial-wave amplitudes of the coupled-channel $\jpsi\pi$ and $D\dvb$ scattering with energy-dependent interaction kernels. 
The perturbative covariant expressions are then unitarized to include the strong final-state interactions. The experimental event distributions of $\jpsi\pi$ and $D\dvb$ measured at 4.23~GeV and 4.26~GeV by the BESIII Collaboration are first separately fitted, and it turns out that the resulting parameters of the scattering amplitudes are quite similar from the two types of fits. 
This implies that the nonperturbative strong interactions of the $\jpsi\pi$ and $D\dvb$ are the responsible mechanism to explain the $\zc$ peaks in our study and the triangle singularities are not necessarily the decisive sources for the observed $\zc$ peaks. 
Based on this observation, we then use the same interaction amplitudes to perform global fits to both the event distributions at 4.23~GeV and 4.26~GeV from BESIII and the lattice finite-volume energy levels. Reasonably good reproduction of the experimental and lattice data is achieved. 

Relevant resonance poles on different Riemann sheets are found for the $\zc$ in our scattering amplitudes. The couplings of the resonance poles with the $D\dvb$ and $\jpsi\pi$ channels are calculated, and the magnitudes of the former channel are found to be around two times larger than those with the latter one. 
The compositeness coefficient of the $D\dvb$, $i.e.$, the probabilities of the $D\dvb$ component in the $\zc$ state, is calculated to be less than 0.5, indicating that other higher-mass hadronic components or compact quark state cores could be also important in the formation of the $\zc$.  It is worthwhile to notice that the $D^*\bar D^*$ channel is only 140~MeV higher than the $D\dvb$ and it can couple to the same quantum numbers as the $\zc$. 
It would be interesting to simultaneously analyze all the data of the $\zc$ and $Z_c(4020)$ in order to understand these charged charmoniumlike structures.


\section*{Acknowledgements}

We thank Mei-Zhu Yan for an early-stage contribution. 
This work is funded in part by the National Natural Science Foundation of China (NSFC) under Grants Nos.~11975090, 12150013, 12047503, 11905258, 12275076, 12125507, 11975075, and 11835015; by the Chinese Academy of Sciences under Grant No. XDB34030000; and by the NSFC and the Deutsche Forschungsgemeinschaft
(DFG) through the funds provided to the TRR110 “Symmetries and the Emergence of Structure in QCD” (NSFC Grant No. 12070131001, DFG Project-ID 196253076). ZHG appreciates the support of Peng Huan-Wu visiting professorship and the hospitality of Institute of Theoretical Physics at Chinese Academy of Sciences, where part of this work has been done.

\section*{Appendix: Covariant partial-wave projection}\label{sec.appendix}
\setcounter{equation}{0}
\def\theequation{A.\arabic{equation}}

One way to describe the angular momenta of a general process $1+2\to \bar{1} + \bar{2}$ is to include the total angular momentum $J$ and its third component $M$, and the orbital angular momentum $\ell$ and the total spin $S$, where only $J$ and $M$ are the good quantum numbers for relativistic reactions. The general partial-wave projection for the process $1+2\to \bar{1} + \bar{2}$ in the $\ell S$ basis is given by~\cite{Gulmez:2016scm} 
\begin{align}\label{eq.gepw}
 V^{J}_{\ell S;\bar{\ell}\bar{S}}(s) =&\, \frac{Y_{\bar{\ell}}^0(\hat{p}_z)}{2(2J+1)} \sum_{\sigma_1,\sigma_2,\bar{\sigma}_1,\bar{\sigma}_2,m} \int d {\hat{\bar{p}}}\, {Y_{\ell}^{m}({ \hat{\bar{p}}})}^{*}\,(\sigma_1 \sigma_2 M| s_1 s_2 S) (m M \bar{M} | \ell S J) (\bar{\sigma}_1 \bar{\sigma}_2 \bar{M}| \bar{s}_1 \bar{s}_2 \bar{S}) \nonumber \\ & (0 \bar{M} \bar{M} | \bar{\ell} \bar{S} J)\,V(p_1,p_2,\bar{p}_1,\bar{p}_2,\sigma_1,\sigma_2,\bar{\sigma}_1,\bar{\sigma}_2)\,,
\end{align}
where the direction of the initial three-momentum $\vec{p}_1$ in the CM frame is defined as the $z$-axis, i.e. $\vec{p}_1=-\vec{p}_2=|\vec{p}|\hat{e}_z$, and the three-momenta of the final-state particles are denoted by $\vec{\bar{p}}_1=-\vec{\bar{p}}_2=|\vec{\bar{p}}|\hat{\bar{p}}$. The energy squared $s$ is given by $s=(p_1+p_2)^2$.  $\sigma_{i}$ and $\bar{\sigma}_{i}$ correspond to the third components of the $i$th particle in the initial and final states, respectively, with $S=\sigma_1+\sigma_2$ and $\bar{S}=\bar{\sigma}_1+\bar{\sigma}_2$. The Clebsch-Gordan coefficient $(m_1m_2m_3|j_1j_2j_3)$ refers to the composition of $\vec{j}_1+\vec{j}_2=\vec{j}_3$, with $m_i$ the third component of $\vec{j}_i$.

For the $S$-wave scattering of the 1 (vector) + 2 (pseudoscalar) $\to$ $\bar{1}$ (vector) + $\bar{2}$ (pseudoscalar) process, the partial-wave projection in Eq.~\eqref{eq.gepw} can be simplified as
\begin{eqnarray}
 V^{J=0}_{01;01}(s) = \frac{1}{2(2J+1)} \sum_{\sigma_1=\bar{\sigma}_1=0,\pm 1} \int d\cos\theta\,\, V(s,t(s,\cos\theta),\sigma_1,\bar{\sigma}_1)\,,
\end{eqnarray}
where $\theta$ denotes the scattering angle defined in the CM frame and the Mandelstam variable $t$ is given by
\begin{eqnarray}
 t= M_{1}^2+M_{\bar{1}}^2- \frac{1}{2s}\left(s+M_{1}^2-M_{2}^2\right)
\left(s+M_{\bar{1}}^2-M_{\bar{2}}^2\right) 
-\frac{\cos\theta}{2s}\sqrt{\lambda(s,M_{1}^2,M_{2}^2)
\lambda(s,M_{\bar{1}}^2,M_{\bar{2}}^2)}\,,
\end{eqnarray}
with $\lambda(a,b,c)=a^2+b^2+c^2-2ab-2bc-2ac$ the K\"all\'{e}n kinematical function. The polarization vectors of the vector meson with mass $m_V$ should be accordingly taken as~\cite{Gulmez:2016scm}
\begin{eqnarray}\label{eq.defeps}
 \varepsilon(\vec{k},0)=  \bma
      \frac{k}{m_V}\cos\theta   \\
      \frac{1}{2}\left(\frac{E_k}{m_V} -1\right)\sin{2\theta}   \\
      0 \\
      \left(1+\cos{2\theta}\right)\frac{E_k}{m_V} -\cos{2\theta}   \\
   \ema\,,\quad 
    \varepsilon(\vec{k},\pm)=  \bma
      \mp\frac{1}{\sqrt{2}}\frac{k}{m_V}\sin\theta   \\
      \mp\frac{1}{\sqrt{2}}\left(\frac{E_k}{m_V}\sin^2\theta + \cos^2\theta \right)   \\
      -\frac{i}{\sqrt{2}} \\
      \mp\frac{1}{2\sqrt{2}}\left(\frac{E_k}{m_V}-1\right)\sin{2\theta}   \\
   \ema\,,
\end{eqnarray}
in order to be consistent with the partial-wave projection formula in Eq.~\eqref{eq.gepw}. 
The magnitude of the three-momentum $k$ and the corresponding energy $E_k$ in Eq.~\eqref{eq.defeps} are defined as $k=|\vec{k}|$ and $E_k=\sqrt{k^2+m_V^2}$, respectively. We explicitly verify that the partial-wave amplitudes obtained from the $\ell S$ basis using Eqs.~\eqref{eq.gepw} and \eqref{eq.defeps} are consistent with the results from the helicity basis~\cite{Jacob:1959at} by using suitable polarization vectors, such as those provided in Refs.~\cite{Lutz:2011xc,Liang:2021fzr}.  Since we only focus on the $S$-wave scattering in this work, the superscript and subscript partial-wave indices $J$ and $\ell S$ always take $J=1$ and $(\ell =0, S=1)$ and are omitted in the main text for simplicity. The explicit expressions after performing the partial-wave projection of the amplitudes in Eqs.~\eqref{eq.vfull22} and \eqref{eq.vfull12} can be found in Eq.~\eqref{eq.vij}.

\bibliographystyle{apsrev4-2-fk}
\bibliography{zc-covariant}

\end{document}